\documentclass[paper]{JHEP3}
\usepackage{epsfig}
\usepackage{graphicx}
\usepackage{xspace}
\usepackage{amssymb}
\usepackage{subfigure}
\usepackage{multirow}

\newcommand{\ra}{\rightarrow}

\newcommand{\be}{\begin{equation}}
\newcommand{\ee}{\end{equation}}

\newcommand{\bea}{\begin{eqnarray}}
\newcommand{\eea}{\end{eqnarray}}
\newcommand{\beanon}{\begin{eqnarray*}}
\newcommand{\eeanon}{\end{eqnarray*}}
\newcommand{\ba}{\begin{array}}
\newcommand{\ea}{\end{array}}
\newcommand{\bd}{\begin{description}}
\newcommand{\ed}{\end{description}}
\newcommand{\bi}{\begin{itemize}}
\newcommand{\ei}{\end{itemize}}
\newcommand{\ben}{\begin{enumerate}}
\newcommand{\een}{\end{enumerate}}
\newcommand{\bc}{\begin{center}}
\newcommand{\ec}{\end{center}}

\newcommand{\GeV}{\mbox{${\mathrm GeV}$}\xspace}

\newcommand{\ordEW}{\mathcal{O}(\alpha_{\scriptscriptstyle EM}^6)\xspace}
\newcommand{\ordEWfour}{\mathcal{O}(\alpha_{\scriptscriptstyle EM}^4)\xspace}
\newcommand{\ordQCD}{\mathcal{O}(\alpha_{\scriptscriptstyle EM}^4
  \alpha_{\scriptscriptstyle S}^2)\xspace}
\newcommand{\ordQCDsq}{\mathcal{O}(\alpha_{\scriptscriptstyle EM}^2
  \alpha_{\scriptscriptstyle S}^4)\xspace}

\newcommand{\eqn}[1]{Eq.(\ref{#1})}

\newcommand{\tbn}[1]{Tab.~\ref{#1}}

\newcommand{\fig}[1]{Fig.~\ref{#1}}

\newcommand{\figsc}[2]{Figs.~\ref{#1},~\ref{#2}}
\newcommand{\sect}[1]{Sect.~\ref{#1}}

\newcommand{\Phantom}{{\tt PHANTOM}\xspace}

\newcommand{\MadEvent}{{\tt MADEVENT}\xspace}

\def\pl #1 #2 #3 {{\it Phys.~Lett.} {\bf#1} (#2) #3}   
\def\np #1 #2 #3 {{\it Nucl.~Phys.} {\bf#1} (#2) #3}
\def\zp #1 #2 #3 {{\it Z.~Phys.} {\bf#1} (#2) #3}
\def\pr #1 #2 #3 {{\it Phys.~Rev.} {\bf#1} (#2) #3}
\def\prep #1 #2 #3 {{\it Phys.~Rep.} {\bf#1} (#2) #3}
\def\prl #1 #2 #3 {{\it Phys.~Rev.~Lett.} {\bf#1} (#2) #3}
\def\intj #1 #2 #3 {{\it Int. J. Mod. Phys.} {\bf#1} (#2) #3}
\def\mpl #1 #2 #3 {{\it Mod.~Phys.~Lett.} {\bf#1} (#2) #3}
\def\rmp #1 #2 #3 {{\it Rev. Mod. Phys.} {\bf#1} (#2) #3}
\def\cpc #1 #2 #3 {{\it Comp. Phys. Commun.} {\bf#1} (#2) #3}
\def\epj #1 #2 #3 {{\it Eur. Phys. J.} {\bf#1} (#2) #3}
\def\jhep #1 #2 #3 {{\it JHEP} {\bf#1} (#2) #3}

\title{Multiple Parton Interactions in $Z+4j$, $W^{\pm} W^{\pm}+0/2j$
and  $W^{+} W^{-}+2j$ production at the LHC}

\author{
Ezio Maina$^{a,b}$\\
$^a$ INFN, Sezione di Torino, Italy,\\
$^b$ Dipartimento di Fisica Teorica, Universit\`a di Torino, Italy
}

\preprint{DFTT 45/2009}

\abstract{
The expected rate for Multiple Parton Interactions (MPI) at the LHC is large.
This requires an estimate of their impact on all measurement foreseen at
the LHC. Conversely it provides new means of studying MPI at the LHC.
In this paper we examine the role of MPI at the LHC,
with the design energy of 14 TeV, in
\bi
\item $Z$ production in association with four jets,
\item $W^{\pm} W^{\pm}$ in association with zero or two jets. 
\item $W^{+} W^{-}$ in association with two jets.
\ei
In all cases the vector bosons are assumed to decay leptonically.

The MPI contribution to $Z+4j$ is dominated by events with two jets with
balancing transverse momentum.
It is possible to achieve a good signal to background ratio,
close to 20\%, for MPI compared to Single
Interaction processes by selecting events with two jets with 
large separation in the transverse plane.
The corresponding statistical significance for
a luminosity of 1 fb$^{-1}$ is about 6.9 for the $\mu^+\mu^- $ channel alone.

The final state channel in which only two same--sign high transverse momentum
charged
leptons are required and additional hard jets are vetoed is dominated by MPI,
with an expected yield of 2500 events with the full LHC luminosity.
}  

\begin{document}

\section{Introduction}
\label{sec:intro}

The presence of Multiple Parton Interactions (MPI) in high energy hadron
collisions has been convincingly demonstrated 
\cite{Akesson:1986iv,CDF_MPI,D0_MPI}.

MPI rates at the LHC are expected to be large,
making it necessary to estimate their contribution to the background of
interesting physics reactions. 
On the other hand, their abundance at the LHC makes it
possible to study MPI experimentally in details, testing and validating the
models which are used in the Monte Carlo's
\cite{Sjostrand:2004pf,Sjostrand:2004ef,Butterworth:1996zw,Bahr:2008dy}
to describe these important features
of hadron scattering.
It is therefore of interest to search for new reactions in
which MPI can be probed and to study in which kinematic regimes they are best
investigated.
Previous studies evaluated the MPI background to
Higgs production in the channel $pp\ra WH \ra  l\nu b \bar{b}$,
\cite{DelFabbro:1999tf}, 
4$b$  production \cite{DelFabbro:2002pw} and $WH$, $ZH$ production
\cite{Hussein:2006xr}.
Recently \cite{Domdey:2009uy} the inclusive double dijet production has been
discussed as a tool to gain information on the two--parton distribution in the
proton. In Ref.~\cite{Calucci:2008jw,Calucci:2009sv} it has been shown how the study
of "inclusive" and "exclusive" multiple interaction cross sections can provide
new information on the non--perturbative structure of the nucleon.

In \cite{Maina:2009vx} MPI have been studied as a background to top--antitop
production at the LHC in the semileptonic channel, particularly
in the early phase of data taking when the full
power of $b$--tagging will not be available.
In the same paper it has been shown that MPI can be accessed in the
$W+4j$ channel, a far more complicated setting than the reactions mentioned before
and that the large cross section for two jet production makes it possible to
detect Triple Parton Interactions (TPI) in $W+4j$ production.

Different reactions involve different combinations of initial
state partons, for instance $\gamma+3j$, $Z+3j$, $W+3j$ MPI processes
test specific sets of quark and gluon distributions inside the proton.
The comparison of several MPI processes will also allow to study the possible
$x$--dependence of these phenomena, namely the dependence
on the fraction of momentum carried by the partons.
CDF found no evidence of 
$x$--dependence in their data which included jets of transverse momentum as low
as five GeV. However in
Ref.\cite{Snigirev:2003cq,Korotkikh:2004bz,Cattaruzza:2005nu} it was shown that
correlations between the value of the double distribution functions for
different values of the two momentum fractions $x_1, x_2$ are to be expected,
even under the assumption of no correlation at some scale $\mu_0$, as a
consequence of the evolution of the distribution functions 
to a different scale $\mu$, which is determined
by an equation analogous to the usual DGLAP equation.
In \cite{Cattaruzza:2005nu} the corrections to the factorized form for the 
double distribution functions have been estimated. They depend on the
factorization scale, being larger at larger scales $Q$, and on the $x$ range,
again being more important at larger momentum fractions. For $Q = M_W$ and
$x \sim 0.1$ the corrections are about 35\% for the gluon-gluon case.    
Moreover  
Ref.\cite{Cattaruzza:2005nu} showed that the correlations in $x_1, x_2$ space
are different for different pairs of partons, pointing to an unavoidable flavour
dependence of the double distribution functions.

In this paper we examine 
\bi
\item the background generated by MPI to $Z+4j\rightarrow \ell^+\ell^- +4j$
production and
the possibility of studying MPI in the $Z+4j$ channel.
\item the observability of MPI in the
$W^{\pm}W^{\pm}\rightarrow \ell^\pm{\ell^\prime}^\pm$ channel.
\item the background generated by MPI to $W^{+}W^{-}+2j$ production and
therefore to Higgs production
via vector fusion in the $H\rightarrow WW \rightarrow \ell\ell\nu\nu$ channel
\ei
at the LHC, with the design energy of 14 TeV.

With its five final state particles, $Z+4j$ production gives the opportunity
to study MPI in a more complex final state than in most 
previous analysis which have typically involved a
combination of two $2 \ra 2$ processes.
The cross section for $Z+4j$ production is expected to be smaller than the
cross section for $W+4j$, mainly because of the smaller branching ratio to
charged leptons in the first case. However the $Z+4j$ channel is cleaner
from an experimental point of view than the
$W+4j$ one since isolated, high pT charged leptons which are the hallmark
of $W$ detection can be copiously produced in B-hadron decays
\cite{iso-lept-from-B}
while no comparable mechanism exists for generating
lepton pairs of mass in the $M_Z$ region.  

The large expected cross section for two jet production suggests that also
Triple Parton Interactions (TPI) could provide a non
negligible contribution in this channel,
as shown to be the case for $W+4j$ processes.

The $W^{\pm}W^{\pm}$ final state has the unique feature that it can be
produced through
MPI at a lower perturbative order, $\ordEWfour$ including W decays, than in
Single Parton Interactions (SPI) which start at $\ordEW$ and $\ordQCD$
with two additional quarks in the final state.
This peculiarity has been noticed before in Ref.~\cite{Stirling_Kulesza_99},
which studied
the inclusive production of two same--sign stable $W$'s at the LHC. Later additional
results concerning the effects of parton correlations have appeared in
the literature\cite{Cattaruzza:2005nu}. Here we treat separately the case
in which the two additional jets are actually observed and the case in which no
jet is required to be present in the final state. In the first case we will
consider all processes contributing to $W^{\pm}W^{\pm}+2j$.
In the second case two different approaches can be adopted: on one hand
the inclusive production of two same--sign $W$'s plus any additional jet activity
can be studied,
on the other hand the focus can be brought to the more exclusive production
of two same--sign $W$'s and no observable jet. In the latter case a jet
threshold is selected and a jet veto is applied: no event with a jet above
threshold is accepted. 

The $W^{+}W^{-}+2j$ channel 
is one of the most important channels for Higgs discovery over a large
portion of the allowed range for the Higgs mass within the SM
\cite{ATLAS-TDR,CMS-TDR} and an estimate of MPI for this final state is
definitely in order.

In \sect{sec:calc} the main features
of the calculation are discussed.
Then we present our results in Sect.~\ref{sec:MPI_in_Z4j}--\ref{sec:MPI_in_WpWm2j}.
Finally we summarize the main points of our discussion.


\section{Calculation}
\label{sec:calc}

The  MPI processes which contribute to $Z+4j$ through Double
Parton Interactions (DPI) are 

\bi
\item $jj \otimes jjZ$
\item $jjj \otimes jZ$
\item $jjjj \otimes Z$.
\ei

For opposite sign $W$'s in $WW+2j$ they are 

\bi
\item $jj \otimes WW$
\item $jW \otimes jW$
\item $jjW \otimes W$
\ei
while for equal sign $W$'s in $WW+2j$ the relevant pairs are

\bi
\item $jW \otimes jW$
\item $jjW \otimes W$
\ei

where the symbol $\otimes$ stands for the combination of one event for each of the two
final states it connects.

The cross section for DPI has been estimated as
\be
\label{eq:sigma_2}
    \sigma =  \sigma_1 \cdot \sigma_2/\sigma_{eff}
\ee
where $\sigma_1 ,\sigma_2$ are the cross sections of the two contributing
reactions.
At the Tevatron, CDF \cite{CDF_MPI}
has measured $\sigma_{eff}=14.5\pm 1.7^{+1.7}_{-2.3}$ mb, a value
confirmed by D0 which quotes
$\sigma_{eff}=15.1\pm 1.9$ mb \cite{D0_MPI}.
In Ref.\cite{Treleani:2007gi} it is argued, on the basis of the simplest two channel
eikonal model for the proton--proton cross section, that a more appropriate value at
$\sqrt{s}= 1.8$ TeV is 10 mb which translates at the LHC into  
$\sigma_{eff}^{LHC}=12$ mb. Treleani then estimates the effect of the removal by CDF
of TPI events from their sample and concludes that CDF measurement yields
$\sigma_{eff} \approx 11$ mb. In the following we conservatively use  $\sigma_{eff}=14.5$ mb
with the understanding that this value is affected by an experimental uncertainty
of about 15\% and that it agrees only within 30\% with the predictions of the eikonal model.
Since $\sigma_{eff}$ appears as an overall factor
in our results it is easy to take into account the smaller value advocated in 
\cite{Treleani:2007gi}.

The only TPI process contributing to $Z+4j$ is

\bi
\item $jj \otimes jj \otimes Z$.
\ei

while the corresponding reaction for $WWjj$, both for opposite and for equal
sign $W$'s production is

\bi
\item $jj \otimes W \otimes W$.
\ei
 
The cross section for TPI, under the same hypotheses which lead to
\eqn{eq:sigma_2}, can be expressed as:
\be
\label{eq:sigma_3}
 \sigma =  \sigma_1 \cdot \sigma_2 \cdot \sigma_3/
 \left( \sigma_{3,eff} \right)^2 /k
\ee

where $k$ is a symmetry factor.
$\sigma_{3,eff}$ has not been measured, and in principle
it could be different from $\sigma_{eff}$. However, in the
absence of actual data, we will assume $\sigma_{3,eff} = \sigma_{eff}$.
In Appendix~\ref{app:A} we present a non rigorous argument which supports the fact
that the two effective cross sections are indeed comparable.
In the following we will keep the TPI contributions, which are
affected by larger
uncertainties, separated from the DPI predictions which are based on firmer
ground.

Three perturbative orders contribute to $4j + \ell^\pm\ell^\mp$ at the LHC
through Single Parton Interactions, while 
two perturbative orders contribute to $\ell\ell^\prime\nu\nu+2j$.
The $\ordEW$ and $\ordQCD$ samples have been generated with \Phantom
\cite{PhantomPaper,method,phact}, while the $\ordQCDsq$ sample
has been produced with \MadEvent \cite{MadeventPaper}.
All reactions contributing to MPI have been generated with \MadEvent .
Both programs generate events 
in the Les Houches Accord File Format \cite{LHAFF}.
In all samples full  matrix elements, without any
production times decay approximation, have been used.
All samples have been generated using CTEQ5L \cite{CTEQ5} 
parton distribution functions.

The relatively high transverse momentum threshold, $p_{T_j} > 30 \mbox{ GeV}$,
and mass separation, $M_{jj}> 60 \mbox{ GeV}$, 
we have adopted for all reactions with jets in the final state
ensures that the processes we are interested in can be
described by (fixed order)
perturbative QCD.

For the $\ordEW$ and $\ordQCD$ samples, generated with \Phantom,
the QCD scale (both in $\alpha_s$ and in the parton distribution functions)
has been taken as

\be
\label{eq:LargeScale}
Q^2 = M_W^2 + \frac{1}{6}\,\sum_{i=1}^6 p_{Ti}^2.
\ee

For the $\ordQCDsq$ sample the scale
has been set to $Q^2 = M_Z^2$. This difference in the scales
leads to a definite relative enhancement of
the  $4j\, +  \, Z$ SPI background and of the MPI contribution compared
to the other ones. Tests in
comparable reactions have shown an increase of about a factor of 1.5 
for the processes computed at $Q^2 = M_Z^2$ with
respect to the same processes computed with the larger scale
\eqn{eq:LargeScale}.
This is the level of uncertainty which is expected for
all the results presented in this paper from variations of the QCD scale.
This estimate is confirmed by the results shown in
Ref.\cite{W+4jNLO:Ellis,W+4jNLO:BlackHat} where the NLO cross section for the
comparable reaction $W+3j$
has been computed and confronted with the LO result.
Other uncertainties stem from the neglect of correlations in the two--particle
distribution functions which, as mentioned in the Introduction, can be as
large as 40\% and from the experimental and theoretical uncertainties on
$\sigma_{eff}$ which range between 15\% and 30\%.
Therefore we expect our prediction to be correct within a factor
of about two.

In order to produce the Multiple Parton Interaction samples we have combined
at random one event from each of the reactions which together
produce the  desired final state through MPI.
When needed, we have required that each
pair of colored partons in the final state have a minimum invariant mass.
This implies that the combined cross section does not in general correspond
to the product of the separate cross sections divided by the appropriate
power of $\sigma_{eff}$ because the requirement of a minimum
invariant mass for all jet pairs induces a reduction of the cross section
when additional pairs are formed 
in superimposing events.

We work at parton level with no showering and hadronization.
Color correlations between the two scatterings have been ignored. They are known
to be important at particle level \cite{Field} but are totally
irrelevant at the generator level we are considering in this paper.


\section{Studying MPI in $Z+4j$ processes}
\label{sec:MPI_in_Z4j}

This reaction shows strong similarities to the $W+4j$ channel studied in
\cite{Maina:2009vx}. In both cases we are dealing with a five body final state and
the MPI cross section is dominated by the $jj \otimes jjV$ mechanism. 
$Z+4j$ rates are smaller than $W+4j$ but the first reaction is somewhat
cleaner from an experimental point of view since leptonically decaying $Z$ can be detected
without ambiguities exploiting the high expected precision for lepton pair masses and are
essentially free of background.

\begin{table}[thb]
\label{Xsection:MPIgencuts}
\vspace{0.15in}
\begin{center}
\begin{tabular}{|c|c|c|}
\hline
Process &  Cross section  & Combined\\
\hline
 $jj$  &  $1.4\times 10^8$ pb & \multirow{2}{20mm}{$3.8\times 10^2$ fb}\\
\cline{1-2}
 $jj\mu^+\mu^-$  &  61 pb & \\
\hline
 $jjj$ &   $7.6\times 10^6$ pb & \multirow{2}{20mm}{\hspace*{5mm} 62 fb} \\
\cline{1-2}
 $j\mu^+\mu^-  $ &   $1.7\times 10^2$ pb & \\
\hline
 $jjjj  $ & $1.2\times 10^6$ pb  & \multirow{2}{20mm}{\hspace*{5mm} 75 fb} \\
\cline{1-2}
 $\mu^+\mu^- $ &   $9.3\times 10^2$ pb & \\
\hline
\end{tabular}
\end{center}
\caption{
Cross sections for the processes which contribute to $4j + \ell^+\ell^-$
through DPI. The selection cuts are given in 
\eqn{eq:cuts}. Notice that the combined cross section corresponds to
$\sigma_1\cdot\sigma_2 / \sigma_{eff}$ only for the $jjjj \otimes Z$ case. In
all other cases there is a reduction due to the requirement of a minimum
invariant mass for all jet pairs since additional pairs are formed 
when the two events are superimposed.
}
\end{table}

\begin{table}[bth]
\label{Xsection:MPIgencuts3}
\vspace{0.15in}
\begin{center}
\begin{tabular}{|c|c|c|}
\hline
Process &  Cross section  & Combined\\
\hline
 $jj$  &  $1.4\times 10^8$ pb & \multirow{3}{13mm}{\hspace*{2mm} 23 fb}\\
\cline{1-2}
 $jj$  &  $1.4\times 10^8$ pb & \\
\cline{1-2}
 $\mu^+\mu^-  $ &   $9.3\times 10^2$ pb & \\
\hline
\end{tabular}
\end{center}
\caption{
Cross sections for the processes which contribute to $4j + \ell^+\ell^-$
through TPI. The selection cuts are given in 
\eqn{eq:cuts}. 
}
\end{table}

In our estimates below we have only taken into account the muon
decay of the $Z$ boson. The $Z\ra e^+e^-$ channel gives the same result.
The possibility of detecting high $p_T$ taus has been
extensively studied in connection with the discovery of a light Higgs in Vector
Boson Fusion in the
$\tau^+\tau^-$ channel \cite{ATLAS-HinVV} with extremely encouraging results.
Efficiencies of order 50\% have been obtained for the hadronic decays of the
$\tau's$. The expected number of events in the $H\rightarrow  \tau\tau
\rightarrow e\mu + X$ is within a factor of two of the yield from 
$H\rightarrow  W W^* \rightarrow e\mu + X$ for $M_H = 120$ \GeV where the
$\tau\tau$ and $W W^*$ branching ratios of the Higgs boson are very close,
suggesting that also in the leptonic decay channels of the taus the efficiency
is quite high. Therefore we expect the $Z \rightarrow \tau^+ \tau^-$ channel
to increase the detectability of the $Z+4j$ final state.

The two jets with the largest and smallest rapidity are identified as forward and
backward jet respectively.
The two intermediate jets will be referred to as central jets in the following.

All samples have been generated with the following cuts:

\bea
\label{eq:cuts}
& p_{T_j} \geq 30~{\rm GeV} \, , \; \; |\eta_j| \leq 5.0 \, , 
\nonumber \\
& p_{T_\ell} \geq 20~{\rm GeV} \, ,\; \;
|\eta_{\ell}| \leq 3.0 \, , \\
& M_{jj} \geq 60 \, {\rm GeV}  \, ,\; \; M_{ll} \geq 20 \, {\rm GeV}\nonumber 
\eea

where $j= u,\bar{u},d,\bar{d},s,\bar{s},c,\bar{c},b,\bar{b},g$.

\begin{table}[h!tb]
\label{Xsection:gencuts}
\vspace{0.15in}
\begin{center}
\begin{tabular}{|c|c|c|c|c|}
\hline
Process &  Cross section &  Cross section &  Cross section &  Cross section\\
\hline
$\ordQCD$  &  $1.1\times 10^2$ fb & 88  fb & 26 fb &  17  fb \\
\hline
$\ordQCDsq$  & $6.4\times 10^3$  fb & $5.6\times 10^3$  fb & 
                                 $2.2\times 10^3$ fb  & $1.4\times 10^3$  fb \\
\hline
$\ordQCDsq_{\mathrm{DPI}}$  & $5.2\times 10^2$  fb & $4.7\times 10^2$ fb & 
                                 $2.7\times 10^2$ fb & $2.5\times 10^2$ fb\\
\hline
$\ordQCDsq_{\mathrm{TPI}}$  & 23  fb & 21  fb & 15 fb  & 15  fb \\
\hline
$\ordEW$  & 17 fb  & 14 fb & 7.6 fb  & 4.8 fb \\
\hline
\end{tabular}
\end{center}
\caption{Cross sections for the processes which contribute to $4j + \mu^+\mu^- $.
For the second column the selection cuts are given in \eqn{eq:cuts}.
For the third column the additional isolation requirement  \eqn{eq:cuts_iso} has been
applied. The events entering the fourth column also satisfy the 
condition \eqn{cuts:DeltaEta} on the separation between the most forward and most
backward jets.
Finally in the last column we present the cross section obtained considering only
events for which the largest azimuthal angular separation satisfies
\eqn{cuts:DeltaPhi}. }
\end{table}

\begin{figure}[htb]
\begin{center}
\mbox{\epsfig{file=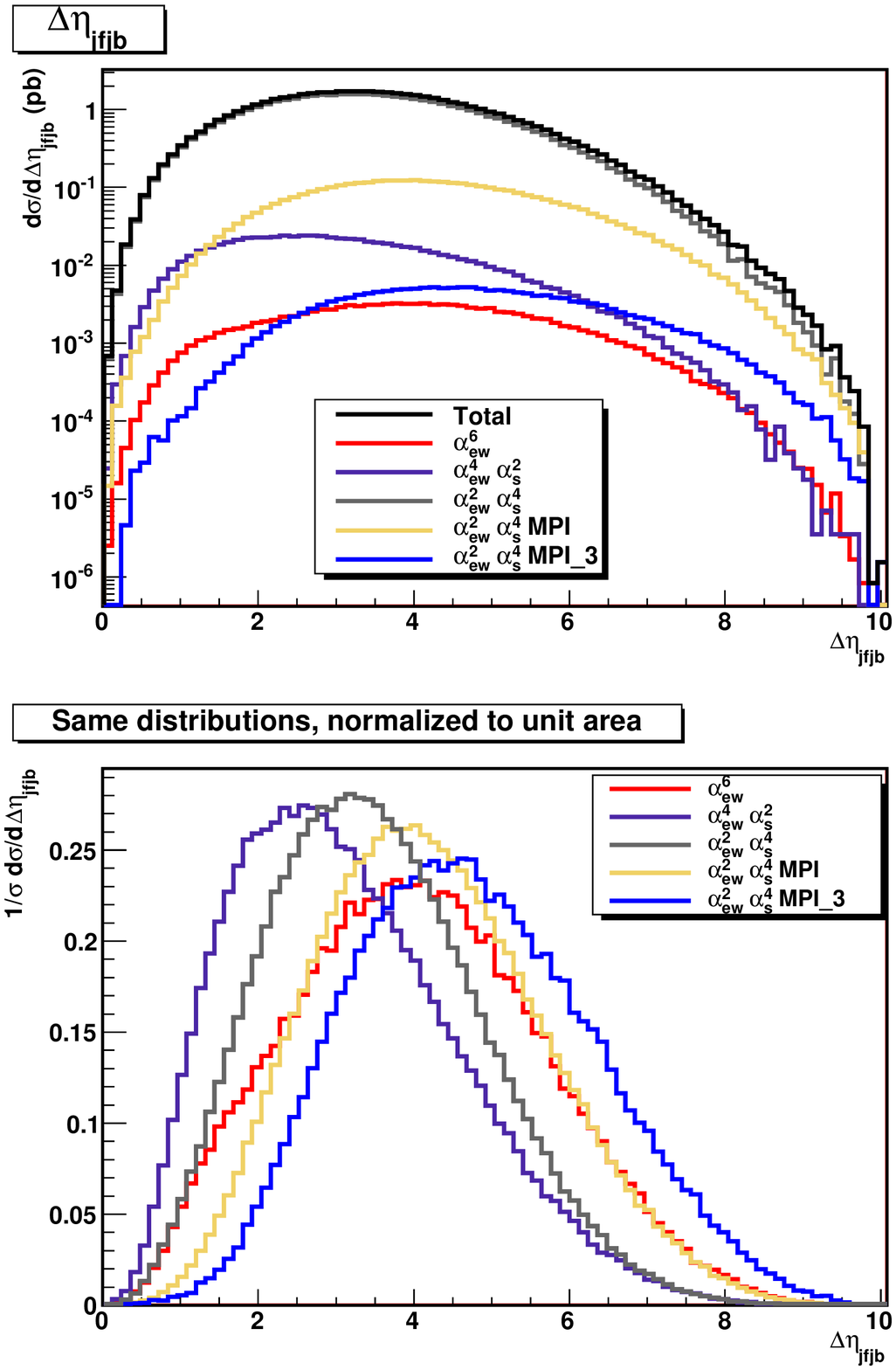,width=10.cm}}
\caption {}
$\Delta\eta$ separation between the most forward and most backward jet 
for the different contributions and for their sum.
Cuts as in \eqn{eq:cuts} and \eqn{eq:cuts_iso}.
The curves in the lower plot are normalized to unit area.
\label{DeltaEta:noflav}
\end{center}
\end{figure}

\begin{figure}[htb]
\begin{center}
\mbox{\epsfig{file=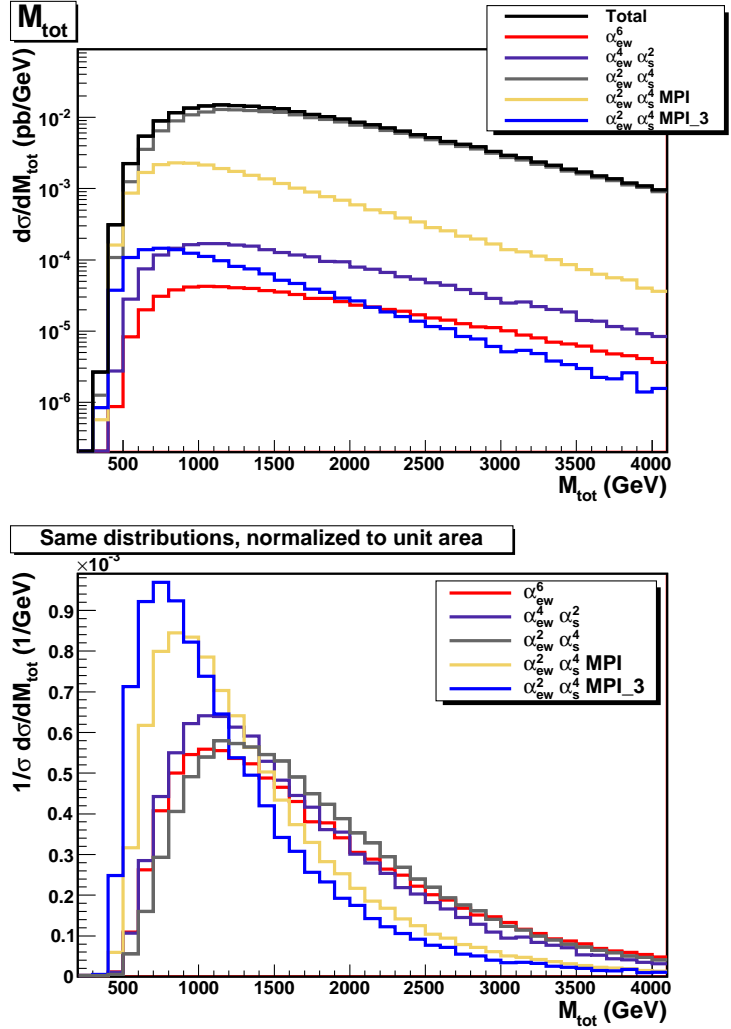,width=10.cm}}
\caption{
Distribution of the total invariant mass of the events
for the different contributions and for their sum.
Cuts as in \eqn{eq:cuts}, \eqn{eq:cuts_iso} and \eqn{cuts:DeltaEta}.
The curves in the lower plot are normalized to unit area.
}
\label{Mvis:noflav}
\end{center}
\end{figure}

\begin{figure}[htb]
\begin{center}
\mbox{\epsfig{file=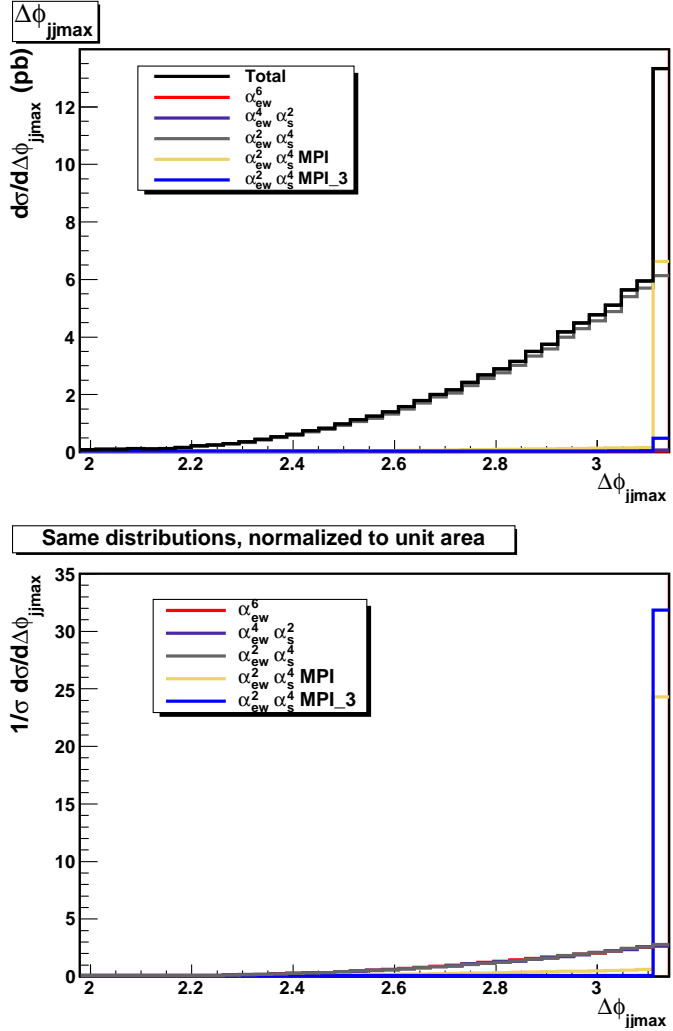,width=10.cm}}
\caption{
Largest $\Delta\phi$ separation between jet pairs
for the different contributions and for their sum.
Cuts as in \eqn{eq:cuts}, \eqn{eq:cuts_iso} and \eqn{cuts:DeltaEta}.
The curves in the lower plot are normalized to unit area.
}
\label{DeltaPhi_jjmax:noflav}
\end{center}
\end{figure}

\begin{figure}[htb]
\begin{center}
\hspace*{-2cm}
\mbox{\epsfig{file=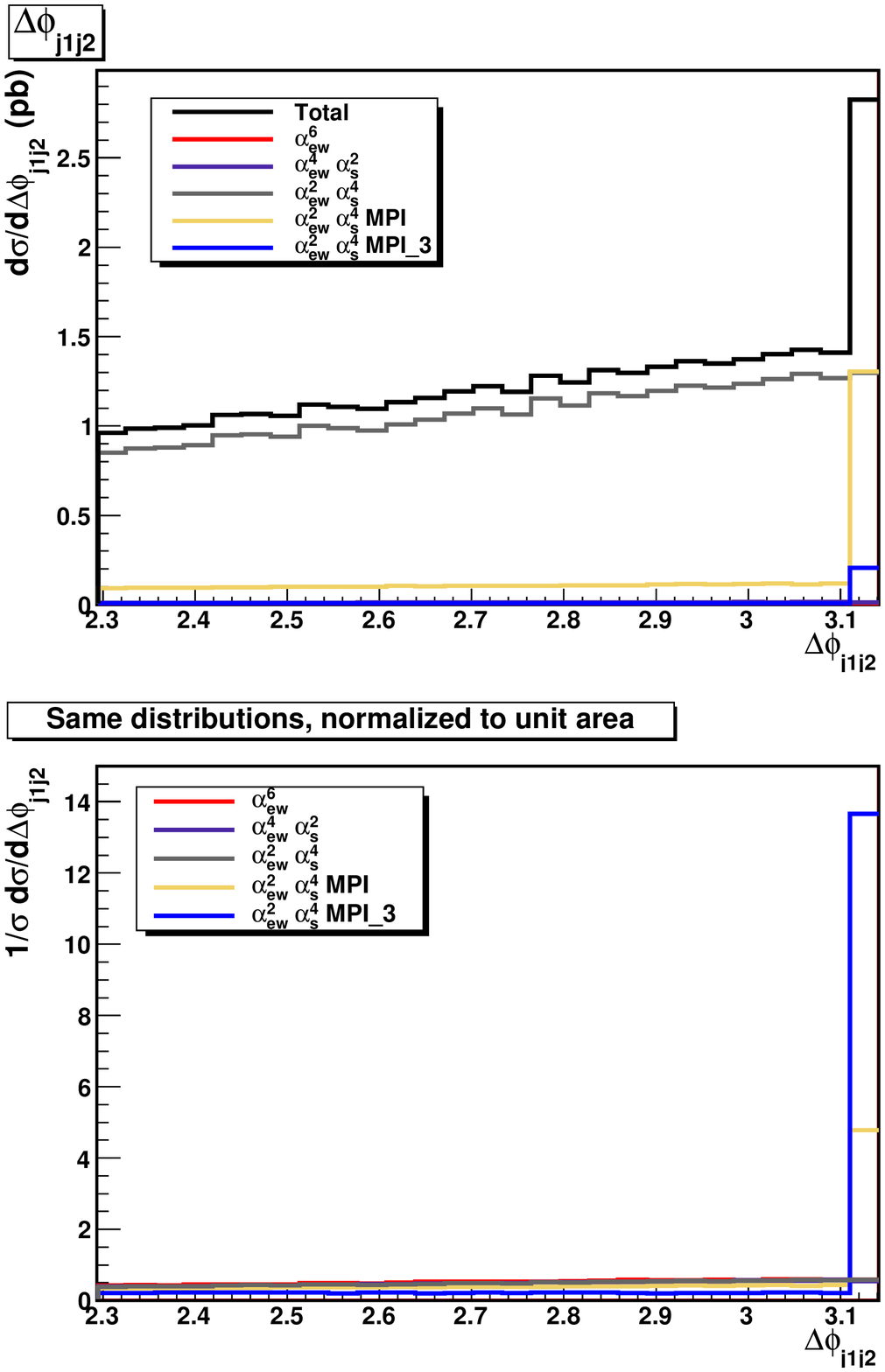,width=8.3cm}}
\hspace*{-0.6cm}
\mbox{\epsfig{file=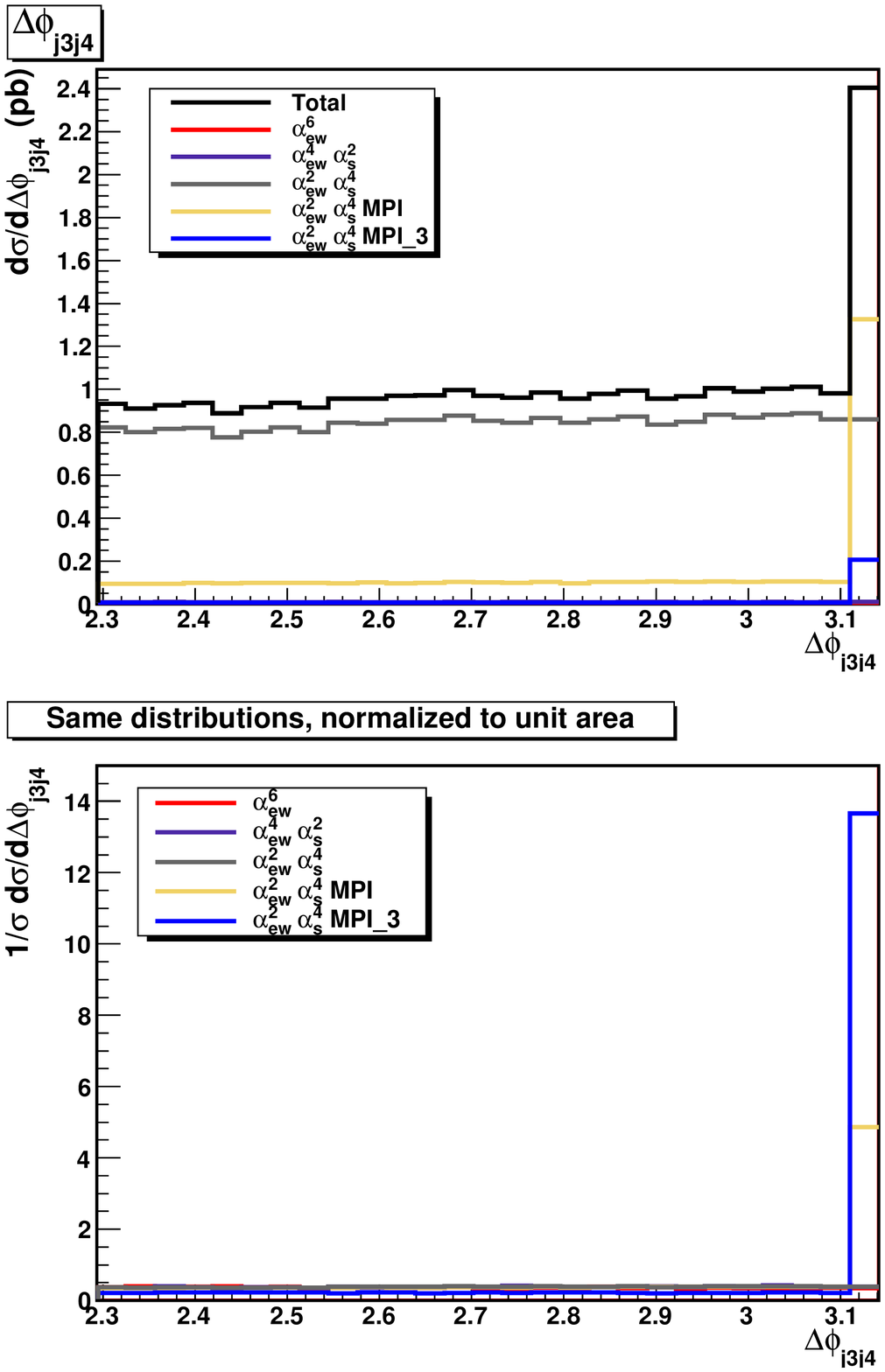,width=8.3cm}}
\hspace*{-2cm}
\caption{
$\Delta\phi$ separation between the two most energetic jets (on the left)
and between the two least energetic among
the four jets (on the right)
for the different contributions and for their sum.
Cuts as in \eqn{eq:cuts}, \eqn{eq:cuts_iso} and \eqn{cuts:DeltaEta}.
The curves in the lower plot are normalized to unit area.
}
\label{DeltaPhi_jj:noflav}
\end{center}
\end{figure}

\allowbreak

The cross sections for the reactions which enter the MPI sample are
shown in
Tab.1 and 
Tab.2
for DPI and TPI respectively.
The largest contribution is given by
processes in which the $Z$ boson is produced in association with two jets
in one interaction and 
other two jets are produced in the second one. As a consequence, as in
the case of $\gamma+3j$ studied by CDF \cite{CDF_MPI} and of the 
$W+4j$ channel most of the events contain
a pair of energetic jets with balancing transverse momentum. The next largest
contribution is due to Drell-Yan processes combined with four jet events.
The smallest, but still sizable, DPI contribution is given by
processes in which the $Z$ boson is produced in association with one jet,
which balances the $Z$ transverse momentum, and the
other three jets are produced in the second interaction.
The cross section for TPI is 23 fb, about 5\% of all MPI processes. 

The cross section for Single Particle Interaction processes and Multiple Parton
Interactions contributing to  the  $jjjj \mu^+\mu^- $ final state,
with the set of cuts in \eqn{eq:cuts}, are shown in the second column of
Tab. 3.
The cross sections in the third column have been obtained with the additional requirements:
\be
\Delta R(jj) > 0.5 \; \;\;\; \;\;
\Delta R(jl^\pm) > 0.5
\label{eq:cuts_iso}
\ee
which ensure that all jet pairs are well separated and that the charged leptons
are isolated from jets.

\fig{DeltaEta:noflav} shows that MPI events tend to have larger separation in
pseudorapidity
between the most forward and most backward jets than $Z+4j$ at $\ordQCDsq$ which
is the only significant background.

Therefore we further require:
\be
\label{cuts:DeltaEta}
|\Delta\eta(j_fj_b)| > 3.8
\ee
In a more realistic environment in which additional jets generated by showering cannot be ignored,
one could impose condition (\ref{cuts:DeltaEta})
on the most forward and most backward of the four most energetic jets
in the event.

The corresponding results are given in the fourth column of 
Tab. 3.
Assuming a luminosity of 1 fb$^{-1}$ this corresponds to a statistical
significance of the MPI $4j + \mu^+\mu^- $ signal of about 6.1 if we take
into account both the
DPI and TPI contributions, and of 5.8 if we conservatively consider only DPI
processes.

\fig{Mvis:noflav} presents the distribution on the invariant mass of the four
jet plus charged leptons system. It shows that typically MPI events are less
energetic than all other contributions considered in this paper.

In \fig{DeltaPhi_jjmax:noflav} we present the distribution of the
largest $\Delta\phi$ separation between all jet pairs.
\fig{DeltaPhi_jjmax:noflav} confirms that MPI processes
leading to $Z+4j$
events are characterized by the presence of two jets which are back to back in
the transverse plane. 
The $Z+4j$ $\ordQCDsq$ SPI contribution displays a much milder increase in
the back to back region. All other contributions are negligible.

The expected $\Delta\phi$ resolution is of the order of
a few degrees for both ATLAS \cite{ATLAS-TDR} and CMS \cite{CMS-TDR} for jets
with transverse energy above 50 GeV. This resolution is comparable
to the width of the bins in \fig{DeltaPhi_jjmax:noflav}.
We have examined the $\Delta\phi$  separation among pairs of jets ordered in
energy, $E_{j_i} > E_{j_{i+1}}$. No clear pattern has emerged.
In \fig{DeltaPhi_jj:noflav} we show the 
$\Delta\phi$ separation between the two most  energetic jets, on the left,
and of the two least energetic ones, on the right.
As might have been guessed by the total mass distribution in
\fig{Mvis:noflav} the ratio between the MPI signal at $\Delta\phi = \pi$ 
and the $Z+4j$ background is somewhat larger for softer jet pairs than for
harder ones. It has proved impossible to clearly associate the two balancing
jets with either the most
forward/backward pair or with the central jets.

We can restrict our attention  to the events for which the maximum  $\Delta\phi$
among jets is in the interval: 

\be
\label{cuts:DeltaPhi}
|\Delta\phi(jj)_{max}| > 0.9\cdot\pi
\ee
The corresponding cross sections are shown in the last column of 
Tab. 3. The rate decrease is of the order of 30\% for Single Parton Interactions
and essentially negligible for MPI processes.  

It appears  quite feasible to achieve a good signal to background ratio,
close to 18/100, for Multiple Interactions  Processes compared to Single
Interaction ones by selecting events with two jets with 
large separation in the transverse plane. The corresponding statistical significance for a 
a luminosity of 1 fb$^{-1}$ is about 6.9 for the $\mu^+\mu^- $ channel alone
with 260/1430 signal/background events. 
\figsc{DeltaPhi_jjmax:noflav}{DeltaPhi_jj:noflav} show that the SPI background is 
smooth in the region $|\Delta\phi(jj)_{max}| \sim \pi$ and almost flat for azimuthal
angular differences among energy--ordered jets while the
MPI signal is mostly concentrated at $|\Delta\phi| \sim \pi$.
This opens the possibility of
measuring the Single Parton Interaction contribution from the neighboring bins
decreasing drastically all theoretical uncertainties on the evaluation
of the background. By measuring ratios of observed events in nearby bins most of the
experimental uncertainties will also cancel. 

\begin{figure}[t]
\begin{center}
\mbox{\epsfig{file=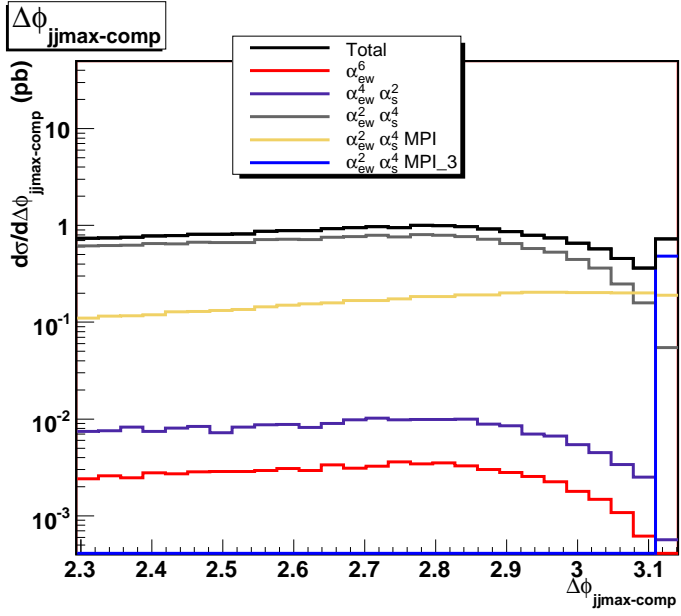,width=10.cm}}
\caption{
$\Delta\phi$ separation between the two jets which do not belong to the pair
with the largest $\Delta\phi$ in the event.
Cuts as in \eqn{eq:cuts}, \eqn{eq:cuts_iso}, \eqn{cuts:DeltaEta}
and \eqn{cuts:DeltaPhi}.
}
\label{DeltaPhi_jjmax_comp}
\end{center}
\end{figure}

\begin{figure}[htb]
\begin{center}
\mbox{\epsfig{file=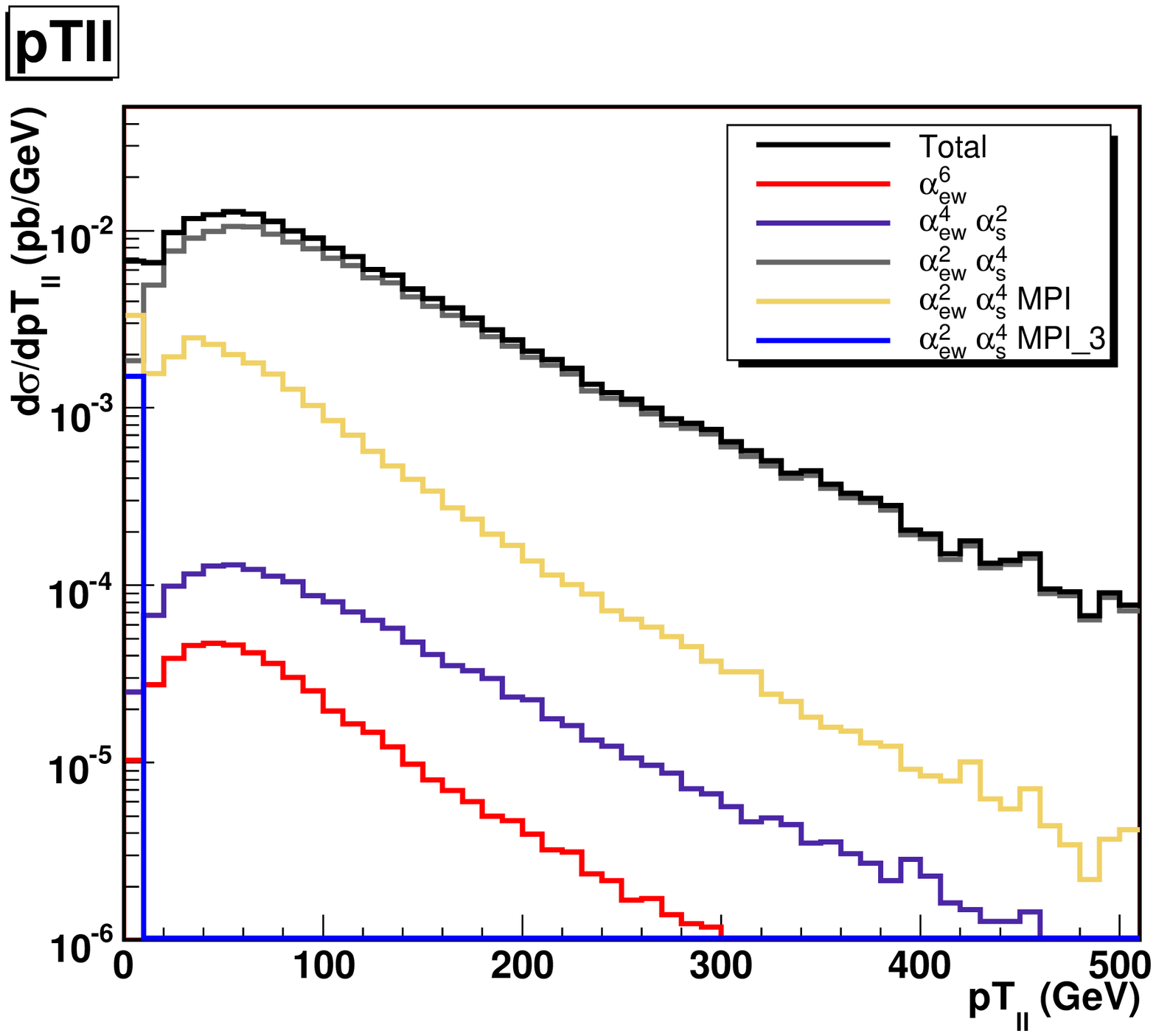,width=10.cm}}
\caption{
Distribution of the transverse momentum of the $l^+l^-$ system.
Cuts as in \eqn{eq:cuts}, \eqn{eq:cuts_iso}, \eqn{cuts:DeltaEta}
and \eqn{cuts:DeltaPhi}.
}
\label{pTll}
\end{center}
\end{figure}

Let us now turn to Triple Parton Interactions in more detail. The obvious
traits which characterize these events are the presence of two pairs of jets
which balance in transverse momentum and of one $Z$ produced by a Drell-Yan
interaction which, to lowest order, has zero the transverse momentum.
While the first feature is not typically found in DPI, $Z$ bosons of Drell-Yan
origin are present in $jjjj \otimes Z$ events which account for about 15\%
of DPI. This is illustrated in \fig{DeltaPhi_jjmax_comp} and \fig{pTll}.
For these two plots we have only considered events satisfying all constraints
in \eqn{eq:cuts}, \eqn{eq:cuts_iso}, \eqn{cuts:DeltaEta}
and \eqn{cuts:DeltaPhi}.

\fig{DeltaPhi_jjmax_comp} shows the angular separation in the transverse plane,
$\Delta\phi_{comp}$,
between the two jets which do not belong to the pair with the largest
$\Delta\phi$ in the event. 
The TPI contribution is concentrated at $\Delta\phi_{comp} \sim \pi$
while all other distributions are rather flat in that region. With the
normalization $\sigma_{3,eff} = \sigma_{eff}$ in \eqn{eq:sigma_3}, 
TPI give  the largest contribution in the bin at $\Delta\phi_{comp} = \pi$,
amounting to more than 50\% of the total. 

\fig{pTll} presents the distribution of the total transverse momentum of the
charged lepton pair; it suggests that the presence of two charged lepton with
essentially zero transverse
momentum is of limited use in separating TPI events
from their background.

The rates for TPI at the LHC are sizable.
Even at low luminosity, L = 30 fb$^{-1}$/year, about 450 TPI events per year are
expected for each charged lepton type.
The corresponding background, integrating over the region
$\Delta\phi_{comp} > 0.9 \pi$, yields about 7500 events leading to a promising
statistical significance larger than five.
Because of the lack of information concerning the rate of Triple Parton
Interactions, it is impossible to draw any firm conclusion from our preliminary
analysis; \fig{DeltaPhi_jjmax_comp} however suggests that indeed it might 
well be
possible to investigate TPI at the LHC exploiting the angular distribution of
pairs of jets with the standard total luminosity expected at the LHC of about
300 fb$^{-1}$ despite the uncertainties which affect the prediction.


\section{Studying MPI in $W^{\pm} W{^\pm} +0/2j$ processes}
\label{sec:MPI_in_WpmWpm2j}

As mentioned in the introduction $W^{\pm} W{^\pm}$ production has the
peculiarity that while the 
SPI contribution starts at $\ordEW$ and $\ordQCD$, the MPI mechanism can produce
two same--sign highly isolated leptons at 
$\ordEWfour$ if no additional jets are required in the final state.

$W^{\pm} W{^\pm}+2j$ production has been shown \cite{CMS:Note_AN2007_005} to be
the vector--vector scattering
reaction which is most sensitive to the details of the EWSB mechanism, which
can be studied in first approximation
comparing cross sections calculated in the presence of a light Higgs and with
the Higgs mass taken to infinity.
Unfortunately the expected rate is small and this channel has to contend with
the contribution to isolated lepton production coming from
B-hadron decays \cite{iso-lept-from-B}.

The inclusive production of same--sign stable $W$'s has been studied in 
Ref.\cite{Stirling_Kulesza_99}, which included all $\ordEWfour$ and 
${\mathcal{O}(\alpha_{\scriptscriptstyle EM}^2 \alpha_{\scriptscriptstyle S}^2)\xspace}$
contributions without taking into account $W$ decays.
In Ref.~\cite{Cattaruzza:2005nu} the effects in this channel of the correlated
evolution of double parton densities have been studied.
While we ignore this issue in the present
analysis, we take into account the decay of the $W$ bosons and require an
experimentally reasonable minimum transverse momentum for the charged leptons.
We also estimate the background due to SM production of
same--sign $W$'s through SPI at $\ordEW$ and $\ordQCD$, including again $W$ decays. 

For $W^{\pm} W{^\pm}+2j$ processes the set of MPI reactions to be included is
the full list mentioned in \sect{sec:calc}.
The corresponding samples have been generated with the set of cuts
shown below:

\bea
\label{eq:cutsWpmWpm2j}
& p_{T_j} \geq 30~{\rm GeV} \, , \; \; |\eta_j| \leq 5.0 \, , 
\nonumber \\
& p_{T_\ell} \geq 20~{\rm GeV} \, ,\; \;
|\eta_{\ell}| \leq 3.0 \, , \\
& M_{jj} \geq 60 \, {\rm GeV}.\nonumber
\eea

If, on the contrary, one aims to reveal MPI production and considers SPI
as a a background then one can resort
to a jet veto in order to suppress the SM SPI contribution.
In this case only the $W \otimes W$ channel has to be
considered in generating the signal.
The additional MPI contributions entering $W W +2j$ production are here
part of higher order
corrections and should be combined with the appropriate virtual contributions
in order to obtain a finite correction.
It is perhaps worth pointing out that the $\ordEW$ and $\ordQCD$
SPI matrix element squared can be integrated over the full phase space without
encountering any soft or collinear singularity.
Therefore, the $\ordEW$ and $\ordQCD$
sample used for the zero--$j$ analysis in this section has
been generated without any constraint
on the final state quarks. The charged leptons are required to satisfy the
standard requirements:

\be
\label{eq:cutsWpmWpm0j}
 p_{T_\ell} \geq 20~{\rm GeV} \, ,\; \;
|\eta_{\ell}| \leq 3.0 \, .
\ee
while no condition is imposed on their combined mass.

\begin{figure}[htb]
\begin{center}
\mbox{\epsfig{file=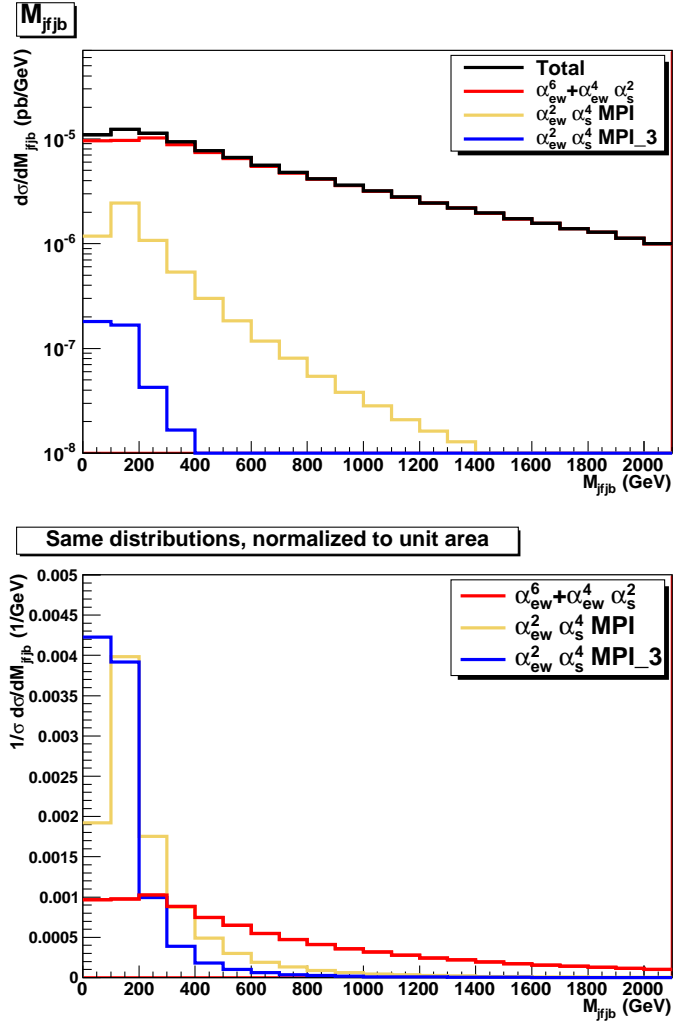,width=10.cm}}
\caption{
Distribution of the invariant mass of the two tag jets in
$W^{\pm} W{^\pm} +2j$ events.
Cuts as in \eqn{eq:cutsWpmWpm2j}. The curves in the lower plot are
normalized to unit area.
}
\label{Mjfjb}
\end{center}
\end{figure}

We will discuss first the case
in which two jets are detected in the final state in
addition to a same--sign lepton pair.  
\fig{Mjfjb} presents the mass distribution of the two tag jets. It shows that
MPI events are concentrated at small
invariant masses while the SPI spectrum extends to very large invariant masses.
Therefore one can improve the
statistical significance of the MPI signal requiring:
 
\be
M_{j_fj_b} \leq 300 ~{\rm GeV}.
\label{eq:cuts_Mjj}
\ee
The cross sections before and after the application of the cut in
\eqn{eq:cuts_Mjj} are given in the second and third
column of \tbn{Xsection:WpmWpm2j} respectively for the $e^{\pm}\mu^{\pm}$
channel which is half of the total same-sign lepton sample.

\begin{table}[h!tb]
\label{Xsection:WpmWpm2j}
\vspace{0.15in}
\begin{center}
\begin{tabular}{|c|c|c|}
\hline
Process &  Cross section &  Cross section \\
\hline
$\ordEW$+$\ordQCD$  &  10  fb & 3.0  fb \\
\hline
$\ordQCD_{\mathrm{DPI}}$  & 0.6  fb & 0.5 fb \\
\hline
$\ordQCD_{\mathrm{TPI}}$  & 0.04  fb & 0.04  fb \\
\hline
\end{tabular}
\end{center}
\caption{Cross sections for the processes which contribute to $W^{\pm}W^{\pm}+2j$
in the $e\mu$ channel.
For the second column the selection cuts are given in \eqn{eq:cutsWpmWpm2j}.
For the third column the additional requirement  \eqn{eq:cuts_Mjj} has been
applied.}
\end{table}

For a luminosity of 300 fb$^{-1}$, which is roughly the total expected luminosity at
the LHC, and taking into account all possible decay
channels to same--sign muons and electrons the statistical significance of the MPI
signal is 6.7 and the expected number of signal/background events is 280/1780
per experiment.
Clearly focusing on relatively soft tag jets makes this result more sensitive
to the presence of additional jets from parton showering.

The final state channel in which only two same--sign high transverse momentum charged
leptons are required has a much larger rate. In the first column of
Table 5
the total cross section with the cuts in \eqn{eq:cutsWpmWpm0j}
are presented, while in the second column we show the results requiring
that no jet with $p_{T_j}\geq 30~{\rm GeV}$ appears in the event.
The ratio between MPI and SPI rates without any jet veto is about 1/3.
The corresponding totally inclusive result presented in Ref.~\cite{Stirling_Kulesza_99}
is appreciably larger, close to 1/2. This difference is due to our cuts on the
charged lepton transverse momentum and pseudorapidity \eqn{eq:cutsWpmWpm0j} which are
more easily satisfied when the two $W$'s are produced in association with two
extra jets and therefore with a non-zero transverse momentum.

\begin{table}[h!tb]
\label{Xsection:WpmWpm0j}
\vspace{0.15in}
\begin{center}
\begin{tabular}{|c|c|c|}
\hline
Process &  Cross section &  Cross section \\
\hline
$\ordEW$+$\ordQCD$ &  14 fb  &  0.9  fb  \\
\hline
$\ordEWfour_{\mathrm{DPI}}$  &  4.3 fb & 4.3  fb \\
\hline
\end{tabular}
\end{center}
\caption{Cross sections for the processes which contribute to $W^{\pm}W^{\pm}+0j$
in the $e\mu$ channel.
The selection cuts are given in \eqn{eq:cutsWpmWpm0j}.
The results in the last column have been obtained vetoing jets with 
$p_{T_j}\geq 30~{\rm GeV}$.}
\end{table}

One sees that only a small fraction of the $\ordEW$+$\ordQCD$ 
events have no hard jet in
the final state and therefore the background is reduced to
only about 20\% of the signal. In the presence of a jet veto the cross section for the
production of two same--sign leptons is dominated by MPI.
The expected rate for all possible combinations of same--sign leptons is about
2500 events per experiment for a luminosity of 300 fb$^{-1}$.
Therefore the $\ordEW$+$\ordQCD$ background is of little concern.
In this case the real
issue are jets faking isolated leptons  and the actual 
isolated leptons from B-hadrons which require a detailed simulation far beyond
the crude estimates presented here.


\section{MPI in $W^{+} W^{-}+2j$ processes: a background to Higgs production
via vector fusion in the $H\rightarrow WW \rightarrow \ell\ell\nu\nu$ channel? }
\label{sec:MPI_in_WpWm2j}

Higgs production in vector boson fusion followed by the decay of the Higgs to a
$W$ pair which in turn decays to two opposite charge leptons and two neutrinos
is arguably the best channel for Higgs discovery over a large portion of the
allowed range for the Higgs mass within the SM \cite{ATLAS-TDR,CMS-TDR}.
In this case no Higgs peak is present in the
data, and more refined analysis are needed. The main background in this channel is
top-antitop production, possibly in association with jets.
In the following we estimate the background provided by $W^{+} W^{-}+2j$
through DPI. The $\ordQCD$ sample includes top-antitop production but misses all 
$t\overline{t}+jets$ processes and as a consequence underestimates the $t\overline{t}$
overall contribution. This is however sufficient since our conclusion is
that the 
DPI $W^{+} W^{-}+2j$ background is overwhelmed by $t\overline{t}$ production.
We roughly follow the analysis scheme 
presented in \cite{CMS:Green_Note_2007_011}.
The contribution from processes in which all external particles are fermions
($8f$), which includes Higgs production as well as all 
$q \overline{q}\rightarrow t \overline{t}$
processes, has been kept separated from the
contribution with two external gluons ($2g6f$), which is 
completely dominated by top-antitop production.
All samples have been generated with the following set of cuts:

\bea
\label{eq:cuts_WpWm2j}
& p_{T_j} \geq 30~{\rm GeV} \, , \; \; |\eta_j| \leq 5.0 \, , 
\nonumber \\
& p_{T_\ell} \geq 20~{\rm GeV} \, ,\; \;
|\eta_{\ell}| \leq 3.0 \, , \\
& M_{jj} \geq 100 \, {\rm GeV}  \, ,\; \; M_{ll} \geq 20 \, {\rm GeV}\nonumber 
\eea
The corresponding cross sections are presented in the second column of 
Table 6.

\begin{table}[bth]
\label{Xsection:WpmWmp2j}
\vspace{0.15in}
\begin{center}
\begin{tabular}{|c|c|c|c|}
\hline
Process &  Cross section &  Cross section &  Cross section \\
\hline
$\ordEW$ + $\ordQCD$ $8f$ &  $9.6\times 10^2$ (2.5) fb &   14 (1.0) fb & 12 (0.9) fb \\
\hline
$\ordQCD$ $2g6f$ &  $6.0\times 10^3$  fb &   26 fb  & 16 fb\\
\hline
$\ordQCD_{\mathrm{DPI}}$  & 5.8  fb &  0.09 fb  & 0.06 fb \\
\hline
$\ordQCD_{\mathrm{TPI}}$  & $2.0\times 10^{-2}$  fb &  $3.0\times 10^{-3}$ fb &  $2.0\times 10^{-3}$ fb \\
\hline
\end{tabular}
\end{center}
\caption{Cross sections for the processes which contribute to $W^{+}W^{-}+2j$.
For the second column the selection cuts are given in \eqn{eq:cuts_WpWm2j}.
For the third column the additional requirement  \eqn{eq:cuts_WpWm2j2} has been
applied. The results in the last column also satisfy \eqn{eq:cuts_WpWm2j3}.
In parentheses, in the fist row, are the cross sections obtained integrating the $8f$
contribution in the mass interval $ 118 \, {\rm GeV}\geq M_{WW} \geq 122 \, {\rm GeV}$
which corresponds in first approximation to the Higgs cross section. 
}
\end{table}

Following Ref.~\cite{CMS:Green_Note_2007_011} we then require the highest
transverse momentum jet to be rather hard and a large separation in pseudorapidity between
the most forward and most backward jets: 
\be
p_{T_{j1}} \geq 50~{\rm GeV}  \, ,\; \; |\Delta\eta(j_fj_b)| > 4.2
\label{eq:cuts_WpWm2j2}
\ee
This leads to the results shown in the third column
of
Table 6.
Finally we require that the two tag jets have a large invariant mass:

\be
M_{jj} \geq 600 \, {\rm GeV}.
\label{eq:cuts_WpWm2j3}
\ee
The corresponding cross sections are shown in the fourth column of 
Table 6.
In parentheses, in the fist row of
Table 6,
are the cross sections obtained integrating the $8f$
contribution in the mass interval $ 118 \, {\rm GeV}\geq M_{WW} \geq 122 \, {\rm GeV}$
which corresponds in first approximation to the Higgs cross section. 

The MPI background is modest to begin with, and is further reduced by the additional
cuts \eqn{eq:cuts_WpWm2j2} and \eqn{eq:cuts_WpWm2j3},
both in absolute terms and in the ratio to the Higgs signal,
to a level at which it can be safely ignored.


\section{Conclusions}
\label{sec:conclusions}

In this paper we have estimated the contribution of
Multiple Parton Interactions to $Z+4j$, $W^{\pm} W^{\pm}+0/2j$ and 
$W^{+} W^{-}+2j$ production.

The MPI contribution to $Z+4j$ is dominated by events with two jets with
balancing transverse momentum.
It is possible to achieve a good signal to background ratio,
close to 20\%, for Multiple Interaction  processes compared to Single
Interaction ones by selecting events with two jets with 
large separation in the transverse plane
and exploiting the expected resolution foreseen by both ATLAS and CMS in the
polar angle $\phi$. The corresponding statistical significance for 
a luminosity of 1 fb$^{-1}$ is about 6.9 for the $\mu^+\mu^- $ channel alone
with 260/1430 signal/background events.
Comparisons with other reactions in which MPI processes can be measured
should allow detailed studies of the flavour and fractional momentum dependence
of Multiple Parton Interactions.
Our preliminary analysis suggests that it might be
possible to investigate TPI at the LHC using the $jj \otimes jj \otimes Z$
channel.

The $W^{\pm} W^{\pm}+2j$ channel has a smaller rate.
For a luminosity of 300 fb$^{-1}$, taking into account all possible decay
channels to same--sign muons and electrons, the statistical significance of the MPI
signal is 6.7 and the expected number of signal/background events is 280/1780
per experiment, with the basic selection  cuts in
\eqn{eq:cutsWpmWpm2j} and \eqn{eq:cuts_Mjj}.

The final state channel in which only two same--sign high transverse momentum charged
leptons are required and additional hard jets are vetoed is dominated by MPI, with an
expected rate of 2500 events with the full LHC luminosity.
The SPI background amounts to about 20\%. Provided the reducible background due to
isolated lepton production in B-hadron decays can be kept under control, 
$W^{\pm} W^{\pm}+0j$ provides a clean opportunity for studying Multiple Parton Interactions
at the LHC.

Finally we have estimated the MPI background to $H\rightarrow WW \rightarrow \ell\ell\nu\nu$
production in the vector fusion channel and found it negligible.

\section *{Acknowledgments}
This work has been supported by MIUR under contract 2006020509\_004 and by the
European Community's Marie-Curie Research 
Training Network under contract MRTN-CT-2006-035505 `Tools and Precision
Calculations for Physics Discoveries at Colliders' 

\newpage
\appendix
\section{A loose argument for the relative size of the effective cross sections in Double
and Triple Parton Interactions}
\label{app:A}

An estimate of the relative size of the effective cross sections $\sigma_{eff}$
and $\sigma_{3,eff}$ for DPI and TPI can be obtained as follows. Let us assume,
being aware that this is  a rather crude approximation,
see Ref.~\cite{Snigirev:2003cq,Korotkikh:2004bz,Cattaruzza:2005nu},
that the
$N$--particle distribution function $\Gamma (x_1,b_1,....,x_N,b_N)$
completely factorizes
\be
\Gamma (x_1,b_1,....,x_N,b_N)=\Gamma_1(x_1,b_1)\cdots\Gamma_N(x_N,b_N).
\ee

Let us also assume that the dependence of two particle distribution function on the momentum fraction
$x$ and on the transverse position $b$ in turn factorize   $\Gamma (x,b) = G(x)f(b)$ where
$G$ is the usual distribution function entering SPI and $f$ is a universal function which does
not depend on the nature of the parton.

We can then write the SPI cross section as:
\bea
\label{eq:sigma_S}
\sigma_S &=& \int G(x_1)\sigma_1(x_1,y_1)G(y_1)\,dx_1 dy_1 \\
         &=& \int G(x_1)f(b_1)\sigma_1(x_1,y_1)G(y_1)f(b_1 - \beta)\, dx_1\, dy_1 \, d^2 b_1\, d^2\beta \nonumber \\
         &=& \sigma_1 \int  T(\beta)\, d^2\beta \nonumber 
\eea
where the overlap function $T = \int f(b)f(b - \beta) \, d^2 b$ takes into account
the dependence on the
impact parameter $\beta$ and on the parton distribution in the transverse plane.
The overlap function, by definition, must be normalized to unity,
$\int  T(\beta) \, d^2\beta = 1$.

Analogously we can write the DPI cross section as follows:

\bea
\label{eq:sigma_D}
\sigma_D &=& \frac{1}{2!} \int G(x_1)f(b_1)\sigma_1(x_1,y_1)G(y_1)f(b_1 - \beta)\,dx_1\, dy_1 \,d^2 b_1 \\
         & &  \qquad G(x_2)f(b_2)\sigma_2(x_2,y_2)G(y_2)f(b_2 - \beta)
         \, dx_2 \, dy_2 \, d^2 b_2 \, d^2\beta \nonumber \\
         &=& \frac{1}{2!} \, \sigma_1 \sigma_2 \int  T^2(\beta) \, d^2\beta \nonumber \\
         &=& \frac{1}{2!} \frac{\sigma_1 \sigma_2}{\sigma_{2,eff}}  \nonumber 
\eea
and in general the N--Parton Interaction cross section can be expressed as:

\bea
\label{eq:sigma_N}
\sigma_N &=& \frac{1}{N!} \int G(x_1)f(b_1)\sigma_1(x_1,y_1)G(y_1)f(b_1 - \beta)
                           \, dx_1 \, dy_1 \, d^2 b_1 \\
         & &  \qquad \qquad \qquad \cdots\cdots \nonumber \\   
         & &  \qquad G(x_N)f(b_N)\sigma_N(x_N,y_N)G(y_N)f(b_N - \beta)
                            \, dx_N \, dy_N \, d^2 b_N \, d^2\beta \nonumber \\
         &=& \frac{1}{N!} \, \sigma_1 \cdots \sigma_N \int  T^N(\beta) \, d^2\beta \nonumber \\
         &=& \frac{1}{N!} \frac{\sigma_1 \cdots \sigma_N}{\sigma_{N,eff}^{N-1}} .\nonumber 
\eea

Therefore 
\label{eq:sigma_eff_N}
\be
\frac{1}{\sigma_{N,eff}^{N-1}} = \int  T^N(\beta) \, d^2\beta . 
\ee

To make progress we can assume for $f$ a simple Gaussian model, which has been extensively
considered in the literature,

\be
\label{eq:f_gauss}
f(b) = \frac{1}{2\pi \, \delta^2} \,  e^{- b^2/(2 \, \delta^2) } .
\ee

In this case
\be
\label{eq:T_N}
\int  T^N(\beta) \, d^2\beta =  \frac{1}{N} \, \frac{1}{(4\pi \, \delta^2)^{N-1}} \, . 
\ee

Therefore the normalization condition is automatically satisfied and
\be
\sigma_{2,eff} = \sigma_{eff} = 2 \, (4\pi \, \delta^2)   
\qquad \sigma_{3,eff} = \sqrt{3} \, (4\pi \, \delta^2) 
 \qquad \sigma_{N,eff} = N^{1/(N-1)} \, (4\pi \, \delta^2)
 \ee
which indeed suggests that all $\sigma_{N,eff}$ are comparable to each other.

\newpage


\begin{thebibliography}{99}

\bibitem{Akesson:1986iv}
T.~Akesson {\it et al.} (Axial Field Spectrometer Collaboration),
\zp C34 1987 163 .

\bibitem{CDF_MPI}
F.~Abe {\it et al.} (CDF Collaboration),\prl 79 1997 584 ,
F.~Abe {\it et al.} (CDF Collaboration), \pr D56 1997 3811

\bibitem{D0_MPI}
The D0 Collaboration, D0 note 5910-CONF.


\bibitem{Sjostrand:2004pf}
T.~Sj\"ostrand and P.Z.~Skands,
\jhep 03 2004 053 , [hep-ph/0402078]. 

\bibitem{Sjostrand:2004ef}
T.~Sj\"ostrand and P.Z.~Skands,
\epj C39 2005 129 , 
[hep-ph/0408302].

\bibitem{Butterworth:1996zw}
J.M.~Butterworth, J.R.~Forshaw and M.H.~Seymour,
\zp C72 1996 637 , [hep-ph/9601371]. 

\bibitem{Bahr:2008dy}
M.~Bahr, S.~Giesekeand M.H.~Seymour,
\jhep 07 2008 076 ,
arXiv:0803.3633 [hep-ph].

\bibitem{DelFabbro:1999tf}
A.~Del Fabbro and D.~Treleani, \pr D61 2000 077502 ,
[hep-ph/9911358].

\bibitem{DelFabbro:2002pw}
A.~Del Fabbro and D.~Treleani, \pr D66 2002 074012 ,
[hep-ph/0207311].

\bibitem{Hussein:2006xr}
M.Y.~Hussein, {\it Nucl.~Phys.~Proc.~Supp.}{\bf B174}(2007)55,
[hep-ph/0610207].

\bibitem{Domdey:2009uy}
S.~Domdey, H.J.~Pirner and U.A.~Wiedemann,
arXiv:0906.4335 [hep-ph].

\bibitem{Calucci:2008jw}
G.~Calucci and D.~Treleani, 
Phys. Rev.D79(2009)034002,
arXiv:0809.4217 [hep-ph].

\bibitem{Calucci:2009sv}
G.~Calucci and D.~Treleani,
Phys. Rev.D79(2009)074013,
arXiv:0901.3089 [hep-ph].

\bibitem{Maina:2009vx}
E.~Maina,
\jhep 04 2009 098 ,
arXiv:0904.2682 [hep-ph].

\bibitem{Snigirev:2003cq}
A.M.~Snigirev,
\pr D68 2003 114012 ,
[hep-ph/0304172].

\bibitem{Korotkikh:2004bz}
V.L.~Korotkikh and A.M.~Snigirev, 
\pl B594 2004 171 ,
[hep-ph/0404155].

\bibitem{Cattaruzza:2005nu}
E.~Cattaruzza, A.~Del Fabbro and D.~Treleani, \pr D72 2005 034022 ,
[hep-ph/0507052].

\bibitem{iso-lept-from-B}
Z.~Sullivan and E.L.~Berger, \pr 74 2006 033008 ,
Z.~Sullivan and E.L.~Berger, \pr 78 2008 034030 .

\bibitem{Stirling_Kulesza_99}
A.~Kulesza and W.J.~Stirling,
\pl B475 2000 168 .

\bibitem{ATLAS-TDR} ATLAS Collaboration, {\it Detector and Physics Performance
Technical Design Report}, 
Vols. 1 and 2, CERN--LHCC--99--14 and CERN--LHCC--99--15.

\bibitem{CMS-TDR} CMS Collaboration,{\it Technical Design Report},  
Vols. 1 and 2, CERN/LHCC 2006--001 and CERN/LHCC 2006--021.

\bibitem{Treleani:2007gi}
D.~Treleani, \pr D76 2007 076006 ,
arXiv:0708.2603 [hep-ph]

\bibitem{PhantomPaper} A.~Ballestrero, A.~Belhouari, G.~Bevilacqua, V.~Kashkan and
E.~Maina, \cpc 180 2009 401 ,
arXiv:0801.3359 [hep-ph]

\bibitem{method} A.~Ballestrero and E.~Maina, \pl B350 1995 225 ,
[hep-ph/9403244].

\bibitem{phact} 
A.~Ballestrero, {\tt PHACT 1.0 - \it Program for Helicity Amplitudes Calculations 
with Tau matrices} [hep-ph/9911318] in {\it 
Proceedings of the 14th International Workshop on High Energy Physics 
and Quantum Field Theory (QFTHEP 99)}, 
B.B.~Levchenko and V.I.~Savrin  eds. (SINP MSU Moscow), pg. 303. 

\bibitem{MadeventPaper}
F.~Maltoni, T.~Stelzer, JHEP 0302 (2003) 027;
T.~Stelzer and W.~F.~Long, Comput. Phys. Commun. {\bf 81} (1994) 357;
J.~Alwall {\it et al.}, arXiv:0706.2334;\\
H. Murayama, I. Watanabe and K. Hagiwara, KEK-91-11.

\bibitem{LHAFF}
J. Alwall {\it et al.},
A Standard format for Les Houches event files.
Written within the framework of the MC4LHC-06 workshop: Monte Carlos for the LHC:
A Workshop on the Tools for LHC Event Simulation (MC4LHC), Geneva, Switzerland, 
17-16 Jul 2005,
\cpc 176 2007 300 ,
[hep-ph/0609017].

\bibitem{CTEQ5}
CTEQ~Coll.(H.L.~Lai {\it et al.}) \epj C12 2000 375 .


\bibitem{W+4jNLO:Ellis}
K.R.~Ellis, K.~Melnikov and G.~Zanderighi,
JHEP 04(2009)077,
arXiv:0901.4101 [hep-ph];\\
K.R.~Ellis, K.~Melnikov and G.~Zanderighi,
arXiv:0906.1445 [hep-ph].

\bibitem{W+4jNLO:BlackHat}
C. FC.~Berger, Z.~Bern, L.J.~Dixon, F.~Febres~Cordero, D.~Forde, T.~Gleisberg,
H.~Ita, D.A.~Kosower and D.~Maitre
Phys. Rev. Lett. 102(2009)222001,
arXiv:0902.2760 [hep-ph];\\
C. FC.~Berger, Z.~Bern, L.J.~Dixon, F.~Febres~Cordero, D.~Forde, T.~Gleisberg,
H.~Ita, D.A.~Kosower and D.~Maitre
arXiv:0907.1984 [hep-ph].

\bibitem{Field}
See for instance the recent talks by R. Field 
http://www.phys.ufl.edu/$\sim$rfield/

\bibitem{ATLAS-HinVV}
S.~Asai {\it et al.}, \epj  C32~S2 2004 19 .

\bibitem{CMS:Note_AN2007_005}
N.~Amapane {\it et al.}, CMS Note AN 2007/005
 
\bibitem{CMS:Green_Note_2007_011}
E.~Yazgan {\it et al.}, CMS NOTE 2007/011.


\end{thebibliography}
\end{document}